\documentclass[runningheads]{llncs}

\usepackage[T1]{fontenc}
\usepackage{amsmath}
\usepackage[noend]{algorithmic}

\usepackage{graphicx}
\usepackage{multirow}
\begin{document}

\title{Federated Learning Approach for Distributed Ransomware Analysis}

%
%
\author{Aldin Vehabovic\inst{1} \and
Hadi Zanddizari\inst{1} \and
Farook Shaikh\inst{1} \and
Nasir Ghani\inst{1} \and
Morteza Safaei Pour\inst{2} \and
Elias Bou-Harb\inst{3} \and
Jorge Crichigno\inst{4} }
%
\authorrunning{A. Vehabovic et al.}
%
\institute{University of South Florida, Tampa, FL 33620, USA \\ \email{\{vehabovica,hadiz,nghani\}@usf.edu} \\ \and
San Diego State University, San Diego, CA 92182, USA \\
\email{msafaeipour@sdsu.edu} \\ \and
University of Texas San Antonio, San Antonio, TX 78249, USA \\ \email{elias.bouharb@utsa.edu} \\ \and
University of South Carolina, Columbia, SC 29208, USA \\ \email{jcrichigno@cec.sc.edu} \\ 
}
\maketitle              
\begin{abstract}
Ransomware is a form of malware that uses encryption methods to prevent legitimate users from accessing their data files. To date, many ransomware families have been released, causing immense damage and financial losses for private users, corporations, and governments. As a result, researchers have proposed a range of ransomware detection schemes using various \textit{machine learning} (ML) methods to analyze binary files and action sequences. However as this threat continues to proliferate, it is becoming increasingly difficult to collect and analyze massive amounts of ransomware executables and trace data at a common site (due to data privacy and scalability concerns). Hence this paper presents a novel \textit{distributed  ransomware analysis} (DRA) solution for detection and attribution using the decentralized \textit{federated learning} (FL) framework. Detailed performance evaluation is then conducted for the case of static analysis with rapid/lightweight feature extraction using an up-to-date ransomware repository. Overall results confirm the effectiveness the FL-based solution.

\end{abstract}
\section{Introduction}
\label{intro}
Ransomware has evolved rapidly over the last decade and is now one of the most serious cyberthreats facing users and organizations today. This malware executes a multi-stage kill-chain to find and infect victim hosts, encrypt their data files, and extract ransom payments.  Expectedly, ransomware represents one of the most lucrative revenue stream for cybercriminals today, many of who now offer \textit{ransomware-as-a-service} (RaaS) \cite{mouss2021} as well. Furthermore, many attackers are also targeting large organizations due to the potential of sizeable payouts. Overall, many different ransomware ``families’’ have been developed and weaponized, with most focusing on Windows users as this is still the most widely-used \textit{operating systems} (OS) type in the enterprise today.

Given these challenges, researchers have been actively studying ransomware analysis schemes in recent years. The main objective in many of these efforts is to identify ransomware programs based on their static or dynamic characteristics and prevent harmful activities. For example, \textit{static analysis} schemes \cite{vehabovic2022} analyze the artifacts of malicious binary files. Hence these methods can be integrated into network-based defenses to analyze incoming files or attachments and detect ransomware earlier in the transmission stage of the kill-chain \cite{mouss2021}. Meanwhile, \textit{dynamic analysis} methods \cite{vehabovic2022} track host system and/or network communication activities to detect ransomware after infection. These schemes can also be integrated into host/network-based defenses, but focus on later stages in the kill-chain (post-infection). Overall, many static and dynamic analysis designs also use \textit{machine learning} (ML) algorithms to process large amounts of ransomware data and extract generalized behaviors, see surveys in \cite{mouss2021},\cite{vehabovic2022},\cite{berrueta2019}.

Nevertheless, as ransomware threats continue to proliferate and diversify, users are being impacted across many domains and sectors. Therefore it is becoming increasingly difficult to collect and analyze massive amounts of ransomware executables and user/system/network trace data at a common (centralized) site. Clearly, this approach poses major privacy and scalability concerns. Foremost, off-site transmission and sharing of sensitive end-host data and network logs is problematic for many users and organizations \cite{vehabovic2022}. Data pre-processing and ML training at a single location also imposes high computational burdens and bandwidth transfer overheads depending upon the type of data being shared (static, dynamic). Inevitably, these limitations complicate the real-world application of many ransomware solutions that use centralized ML training. In addition, most studies have used datasets containing a mixture of older ransomware families targeting Windows 7/8 systems (from mid-2010s time frame) \cite{vehabovic2022}. 

In light of the above, there is a pressing need to develop new ransomware analysis schemes to detect \textit{and} attribute the latest threats with improved scalability and privacy. Of key concern are threats targeting Windows 10/11, the most prevalent OS in enterprise environments. Hence this paper presents a novel ransomware solution which leverages \textit{federated learning} (FL), a decentralized and collaborative ML framework designed to address scalability and privacy concerns \cite{mcmahan2017},\cite{qli2021}. Indeed, FL has already been applied to range of problems in image/voice recognition, keystroke prediction, smart grids, fraud detection, etc \cite{qli2021},\cite{hudson2021}. Recent studies have also proposed some FL-based cybersecurity schemes, e.g., for \textit{Internet of Things} (IoT) intrusion detection, network anomaly detection, and traffic recognition \cite{zhao2019},\cite{li2021},\cite{nguyen2019},\cite{rey2021}.  However the further application of this distributed ML paradigm for ransomware analysis (particularly detection and attribution) has not yet been considered. Hence this topic forms the key motivation for the work herein, and several key contributions are made:
\begin{enumerate}
\item Novel \textit{distributed ransomware analysis} (DRA) architecture for ransomware detection and attribution based upon the FL framework. This solution embodies a generic approach which can implement a wide range of static and dynamic ransomware analysis schemes.
\item Use of a new ransomware dataset repository with some of the latest families targeting Windows 10/11, i.e., including Babuk/Babyk, BlackCat, Chaos, DJVu/STOP, Hive, LockBit, Netwalker, Sodinokibi/REvil, and WannaCry. Since new malware can have much fewer available samples, this repository is limited to under 1,500 binary executables to model realistic scenarios. 
\item Detailed evaluation of the DRA framework for the case of static ransomware analysis.  Specifically, rapid/lightweight feature extraction is done using Windows \textit{portable executable} (PE) files, and FL performance is then tested using decentralized \textit{neural-network} (NN) based classifiers. 
\end{enumerate}
To the best of the authors' knowledge, this is the first known study on FL for ransomware detection and attribution. This paper is organized as follows. Section \ref{survey} presents a review of ransomware detection schemes as well as recent FL-based cybersecurity schemes. Subsequently, Section \ref{architecture} details the new DRA architecture and its distributed FL algorithm. The ransomware dataset repository is then detailed in Section \ref{dataset}, including sample collection and feature selection/extraction. Detailed performance results are then presented in Section \ref{performance} for the case of static analysis, followed by conclusions and future work directions in Section \ref{conclusions}.

\section{Literature Review}
\label{survey}
Ransomware follows a well-established kill-chain sequence comprised of several key stages, i.e., reconnaissance, distribution/delivery, installation/infection, communication, encryption, and extortion/payment \cite{vehabovic2022}. As a result, researchers have proposed a range of schemes for ransomware detection and mitigation targeting different stages in this sequence. A brief taxonomy and overview of some of these methods is presented along with recent FL-based security solutions. 

\subsection{Ransomware Detection}
\label{survey_ransomware}
%

Ransomware analysis has received much attention in the past decade. Various survey articles have also appeared in this domain, detailing different (sometimes overlapping) taxonomies to classify the proposed solutions, e.g., including static or dynamic analysis schemes, network-based or host-based methods, forensic analysis techniques, etc \cite{mouss2021}-\cite{berrueta2019}.  Consider some details.

Static analysis methods analyze binary malware executables to detect artifacts of malicious behaviors \cite{mouss2021}. Common techniques here include code analysis for malware author attribution, code/segment de-anonymization, reverse engineering, ransomware server address and domain prediction, etc \cite{vehabovic2022}. For example, \cite{poudyal2018} specifies a multi-level framework to classify ransomware. This scheme analyzes raw binary files, assembly code, and libraries using Linux object-code dumps and portable executable parsers. ML classifiers are then trained using the extracted data, yielding detection rates around 90\%. Meanwhile, \cite{zhang2019} presents a scheme to analyze operational code sequences.  These sequences are transformed into N-grams, and \textit{term frequency-inverse document frequencies} (TF-IDF) features are generated to train classifiers (decision tree, RF, etc). Results show detection rates around 90\%. However, code-based analysis is a slow and labor-intensive approach \cite{Mulders2017} which is better suited for post-infection forensic analysis. 

Recent efforts have also generated other features for static analysis. For example, \cite{wang2021} presents a unique ransomware classification scheme which uses image processing for feature extraction. Namely, ransomware binary files are first converted to grayscale images, and texture analysis is then used to compute features. Results for several classifiers show good accuracy (97\%) for a small dataset containing a mix of older and newer ransomwares (379 samples). However, this scheme imposes higher computational burdens for large datasets and does not consider benign applications. Meanwhile, \cite{zhu2022} details another static ransomware analysis scheme which computes/extracts entropy and image-based features from binary files and uses them to train a Siamese NN classifier. Tests for a small dataset with about 1,000 samples and 10 ransomware families show accuracy values in the mid-90\% range but notably lower precision and recall rates (upper 70\% range). Also, most of the ransomware families used here are older (from mid-2010s time frame) and benign applications are not considered.

Researchers have also used Windows PE format files for static malware analysis. These files contain metadata and other information for binary executables and enable very fast/lightweight feature extraction, i.e., versus more compute-intensive image or entropy-based methods above. However, most studies using PE files have focused on the generic malware detection problem (and not ransomware classification). For example, a 2016 study \cite{kim2016} uses malware samples from {\tt VX Heaven} (now inactive) to train several ML-based detection classifiers using about 10 PE file features. Results show detection rates over 90\%. Meanwhile, \cite{liao2021} extracts PE file features from a dataset with 5,500 malware and 1,200 benign samples (from the early 2010s) and uses customized rules to achieve 95\% detection rates. Also, \cite{rezaei2020} extracts a small set of PE file features from a dataset with 1,200 malicious and benign samples. Several classifiers are then trained, yielding over 95\% detection accuracy. However, these studies use malware datasets which are almost a decade old and provide few details on composition. To address this concern, \cite{vehabovic2023} presents an updated study on ransomware detection \textit{and} attribution using static PE file analysis. A new repository is curated comprising of 9 of the latest ransomware families (about 1,200 samples) and benign applications (2,000 samples). 
 Results for several ML classifiers show impressive detection and attribution percentages in the mid-90\% range with up to 15 features.

Meanwhile dynamic analysis examines run-time actions and events at the network and/or host levels. Namely, dynamic network-based schemes analyze packet traces for ransomware activity, e.g., server communications, \textit{domain name service} (DNS) queries, networked storage access, etc. For example, \cite{almash2019} presents a detection scheme for the Locky ransomware (2016) which collects behavioral/non-behavioral traffic features in a testbed and then trains several classifiers. Results show a mean detection rate of 97\%.  
Meanwhile, \cite{homayoun2019} presents a ransomware detection scheme using advanced \textit{neural network} (NN) classifiers which yields over 97\% detection rates for older ransomware threats such as CryptoWall, TorrentLocker, and Sage. The NetConverse scheme \cite{alhawi2018} also uses ML methods to analyze Windows host traffic for several families from the 2010s time frame. Results show high detection rates, over 95\%. Finally, \cite{roy2021} presents a deep learning approach to detect and classify abnormal traffic from Windows 7 hosts, and results show high detection rates for several families.

Dynamic host-based schemes monitor local system activities to detect ransomware operations and possibly recover encryption keys. These methods are more latent as they use virtual run-time environments to execute binary files and capture traces, i.e., \textit{virtual machines} (VM) or sandboxes. A range of actions can be tracked here, including \textit{application programmer interface} (API) calls, \textit{dynamic link library} (DLL) calls, and memory and file operations. For example, \cite{kharraz2016} monitors for file encryption/deletion, persistent desktop messages, etc. Results show successful detection of about 96\% of older ransomware. Meanwhile, \cite{kolodenker2017} presents a scheme to monitor/store encryption keys and facilitate ransomware detection and recovery. This solution can mitigate about 12 out of the 20 families tested. 
Recent efforts have also targeted ransomware ``paranoia'' activities, i.e., pre-attack actions to detect environments and avoid fingerprinting. For example, \cite{molina2021} monitors pre-attack API calls and uses \textit{natural language processing} (NLP) to extract features. Findings confirm that many ransomware families generate distinguishable API fingerprints. 
Also, \cite{molina2022} presents a host-based framework for API call monitoring. Some older ransomware families (such as Reveton, Locky, Teslacrypt, etc) are fingerprinted to extract features and build frequency pattern trees for real-time detection of API sequences.

Although the above works provide some key contributions, notable concerns still remain. Foremost, the majority of studies have focused on older ransomware families targeting dated Windows 7/8 systems (mid-2010s). Given the persistent nature of this malware, it is imperative to  focus on newer threats to Windows 10/11 OS users. Furthermore, new releases will likely have fewer samples to analyze, posing added challenges. Hence ``data-centric'' ML solutions must achieve effective performance with constrained  datasets. Finally, ransomware detection/attribution schemes must have amenable run-time complexity and ideally tackle threats in the earlier distribution/delivery stages to minimize damage \cite{vehabovic2022}. It is here that static analysis is more expedient for examining suspicious executable payloads/downloads. By contrast, dynamic analysis requires more latent tracking and analysis of network or host activities and is is better suited for latter stages in the kill-chain. Finally, it is important to consider attribution, i.e., ransomware classification, as this represents a logical next step after detection.

\subsection{Federated Learning (FL) in Cybersecurity}
\label{survey_FL}
FL is a decentralized learning framework developed by Google in 2016. This solution uses multiple end-point clients to train ML models under the coordination of a central server \cite{mcmahan2017}.  Namely, ``global'' learning is done over multiple (synchronous or asynchronous) communication rounds. In each round, the server selects a subset of clients and sends them its latest global ML model parameters. Clients then perform ``local'' training with their own data and only send back model parameter updates. The central server uses these updates to revise its global ML model parameters and then repeats the process. 

Overall, FL provides some key benefits for large ML problems. Foremost, user privacy is greatly improved as sensitive data stays at local hosts.  Computational scalability is also much better since many systems are involved in the distributed learning process. Finally, bandwidth scalability (efficiency) also increases since raw datasets do not need to be transferred to a central server. 
As a result, FL has seen strong traction in a diverse range of areas, e.g., image/voice recognition, keystroke prediction, smart grid operation, fraud detection etc \cite{qli2021},\cite{hudson2021}. Many further variants of this approach have also been proposed, as surveyed  in \cite{kairouz2021}.  Although a detailed review of these works is out of scope herein, some key contributions include revised training and averaging methods for unbalanced datasets, compression/quantization methods to reduce FL communication overheads, client failure recovery techniques, detection and mitigation of malicious clients (adversarial FL), and fully distributed ``peer-to-peer'' FL schemes.

Now researchers have also applied FL to the security domain. For example, \cite{zhao2019} presents a FL framework to train \textit{deep NN} (DNN) models for multiple tasks such as anomaly detection, traffic recognition, and classification. Tests with older traffic profiles (2016) show good detection accuracies for various anomalous events, on par with centralized schemes. Meanwhile, \cite{li2021} details a FL-based DeepFeed solution for intrusion detection in cyberphysical systems using \textit{convolutional NN} (CNN) models. Results with real-world industrial network data show good detection accuracy, exceeding some centralized schemes. Meanwhile, \cite{nguyen2019} presents a FL scheme for anomaly detection of compromised IoT devices. Here, local access gateways monitor traffic and map packet streams to symbols. Language processing methods are then used in conjunction with \textit{recurrent NN} (RNN) algorithms. Results for the Mirai malware show very good attack detection (over 95\%) and low run-time complexity. Finally, \cite{rey2021} presents another FL approach for IoT malware detection using supervised and unsupervised ML algorithms for anomaly detection (with packet data). Again, FL gives very good performance, closely matching centralized classifiers. The authors also test several adversarial FL setups with model poisoning attacks. Results show a notable decline in accuracy when over 25\% of clients are compromised. However, as noted earlier, the further application of FL for ransomware analysis has not been considered.

\section{Distributed Ransomware Analysis (DRA)}
\label{architecture}
The continued growth of ransomware is posing many operational challenges for threat detection and mitigation. Most notably, it is increasingly difficult (infeasible) to collect large amounts of raw data from massive user bases and transfer it to a common site for analysis. Indeed, privacy concerns will prevent many users and organizations from sharing their sensitive file/trace logs in the first place.  Additionally, bandwidth transfer overheads can be very high, and a single centralized computing facility may not be able to process extreme amounts of data and train large models. Inevitably, these constraints will limit the effectiveness of ML-based solutions. Hence there is a pressing need to develop new ransomware analysis frameworks which address these critical concerns.

\begin{figure*}[h]
    \centering
    \includegraphics[width=5.0in, height=2.5in]{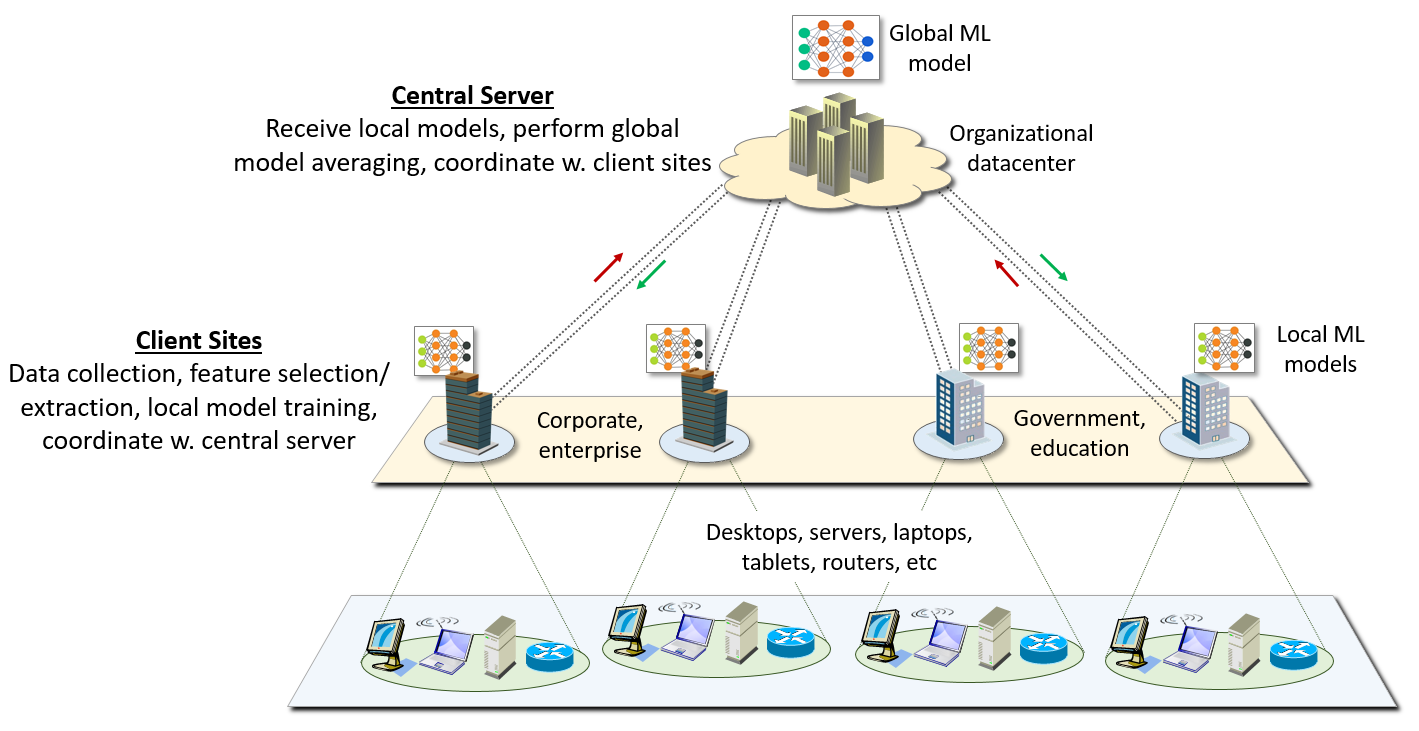}
    \caption{Distributed ransomware analysis (DRA) framework using federated learning}
    \label{DRA_architecture}
\end{figure*}

Now scalability and privacy demands in large ML problems are not new, and researchers have proposed various solutions such as privacy preserving computation and \textit{federated learning} (FL) \cite{mcmahan2017} (Section \ref{survey_FL}).  Of these, the latter is the most scalable as it distributes and decentralizes training over multiple computational nodes. Hence a novel \textit{distributed ransomware analysis} (DRA) solution is developed using the FL framework. This setup is shown in Fig. \ref{DRA_architecture} and comprises of several \textit{client sites} communicating with a \textit{central server}. This architecture features a generic design that can be tailored for a full range of ransomware analysis schemes, both static and dynamic (Section \ref{survey_ransomware}). Furthermore, it is assumed that the whole setup operates in a trusted manner, i.e., careful vetting/pre-selection of client sites, authenticated and encrypted communications, encryption of training data, etc. The DRA architecture and its associated FL algorithm are now presented, followed by a detailed performance evaluation study in Section \ref{performance}.

\subsection{Client Sites}
\label{architecture_client}
Client sites in the DRA architecture mirror the roles of client nodes in the FL architecture \cite{mcmahan2017}. These entities are located at carefully-selected, trusted organizations with large user bases, e.g., such as corporate campuses, government facilities, academic institutions, and even network service provider \textit{points of presence} (PoP). Note that it is not  feasible for individual systems (such as servers, laptops, etc) to act as FL clients. The main reason here is that individual devices will not be able to collect a sufficient number of ransomware samples for effective ML training (data scarcity). Hence DRA client roles are placed at vetted institutional sites which have access to much larger volumes of raw data from many users and systems. As there are generally fewer client sites involved here, i.e., tens-hundreds, this approach embodies a cross-silo FL approach \cite{rey2021}.
\begin{figure*}[h]
    \centering
    \includegraphics[width=4.75in, height=2.35in]{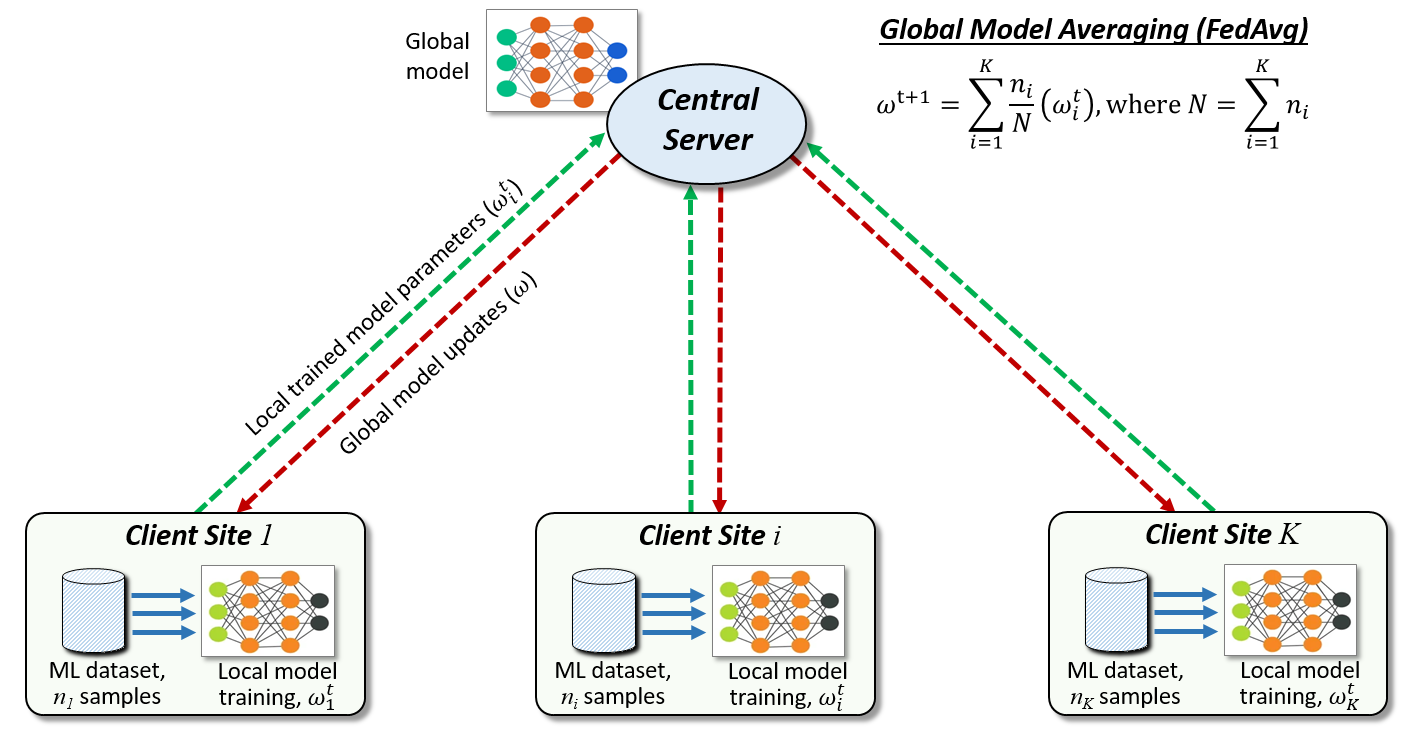}
    \caption{Logical view of federated learning (FL) algorithm in DRA architecture}
    \label{FL_alg}
\end{figure*}

Overall, DRA client sites perform several key functions, including raw data collection, feature selection/extraction, local ML model training, and communication with the central server.  Foremost, raw data samples are collected from the internal user base and other systems, e.g., desktops, laptops, tablets, mail and web servers, routers, and even cyberphysical devices. Depending upon the ransomware analysis approach being implemented (Section \ref{survey_ransomware}), data acquisition can entail a wide range of possibilities, e.g., such as extracting email files from mail servers, capturing webpage attachments from servers, logging router packet traces, extracting host/system call logs, etc. Client sites then perform further data pre-processing and feature selection/extraction on the raw data samples. Namely, feature selection identifies a subset of parameters of interest, whereas feature extraction selects parameters to generate ML insights. The extracted features are then collected to build a local ML training/testing dataset, Fig. \ref{DRA_architecture}. 

\begin{figure}[!ht]
\line(1,0){350}
\begin{algorithmic}
\small
\STATE \textbf{Input:} Receive global ML model parameters from central server, $\omega$ 
\\
\vspace{0.05in}
\STATE Partition local ML dataset ($n_i$ samples) into $B$ batches
\\
\vspace{0.05in}
\STATE \textit{/* Train local ML model over given epochs and batches */}
\FOR{j=1 to $E$}
\FOR{k=1 to $B$}
  \STATE Train local model over $k$-th batch of data
  \STATE Update local ML model parameters $\omega \rightarrow \omega_i^t$
\ENDFOR
\ENDFOR

\vspace{0.05in}
\STATE \textbf{Output:} Send updated ML model parameters, $\omega_i^t$, to central server
\end{algorithmic}
\line(1,0){350}
\caption{Centralized FL local training algorithm at client site $i$}
\label{psuedocode_client}
\end{figure}

Now client sites must also communicate with the central server to participate in the distributed FL training process. A logical overview of this strategy is shown in Fig. \ref{FL_alg}, along with sample client site psuedocode in Fig. \ref{psuedocode_client}. Namely, the training process is jumpstarted when a client site receives parameters for the initial global ML model from the central server, i.e., denoted by $\omega$. Since NN-based algorithms are most amenable to FL implementation \cite{qli2021}, these parameters will typically correspond to weight vectors for \textit{feedforward NN} (FNN), \textit{deep NN} (DNN), \textit{convolutional NN} (CNN), or \textit{long short-term memory} (LSTM) networks \cite{geron2019}. Client sites then train the received model using their local data. In particular, it is assumed that client site $i$ has a local ML dataset with $n_i$ samples, and this is further split into $B$ batches for training over $E$ epochs, see Fig. \ref{psuedocode_client}. Finally, the updated local model parameters at each round $t$, $\omega_i^t$, are sent back to the central server for further processing and updating. Overall, localized model training allows client sites to retain their sensitive data.

\subsection{Central Server}
\label{architecture_central}
The central server in the DRA architecture manages the distributed, decentralized ML training process for ransomware analysis. Akin to its namesake in the FL setup \cite{mcmahan2017}, this entity communicates with the client sites to update the global ML model parameters. Sample psuedocode for the central server algorithm is also presented in Figure \ref{psuedocode_server} based upon the {\tt FedAvg} algorithm \cite{mcmahan2017}. Namely, training is done over $T$ communication rounds with the client sites. In each round, the central server selects a subset of $K$ client sites to participate in the training and sends them its latest global model parameters, $\omega$. It then waits for the local client training sessions to complete and receives/processes parameter updates, i.e., where $\omega_i^t$ denotes the updated ML model parameters from client site $i$ in round $t$ (Figs. \ref{psuedocode_client}, \ref{psuedocode_server}). Note that Fig. \ref{psuedocode_server} shows a synchronous FL training approach where all clients respond in each iteration. However, this detail is not specific to the DRA architecture and can be easily modified for asynchronous operation, e.g., by specifying client site selection and update processing \cite{mcmahan2017},\cite{qli2021}.

\begin{figure}[!ht]
\line(1,0){350}
\begin{algorithmic}
\small
\STATE  \textbf{Input:} Initial global ML model parameters, $\omega_0$ 
\\
\vspace{0.05in}
\STATE \textit{/* Iterative FL training process over $T$ rounds */} 
\FOR {$t=1$ to $T$}

\vspace{0.05in}
\STATE Select $K$ client sites 
\\
\vspace{0.05in}
\FOR {$i=1$ to $K$}
\STATE Send latest global ML model $\omega$ to $i$-th client site
\ENDFOR

\vspace{0.05in}
\STATE Wait to receive local updates from all $K$ client sites, $\omega_i^t$ 
\\
\vspace{0.025in}
\STATE Average to update global ML model parameters: $\omega  \leftarrow \omega^{t+1}=\sum_{i=1}^{K} \frac{n_i}{N} \omega_i^t$
\ENDFOR

\vspace{0.05in}
\STATE \textbf{Output:} Final global ML model parameters, $\omega$ \\
\end{algorithmic}
\line(1,0){350}
\caption{Centralized FL averaging algorithm at central server}
\label{psuedocode_server}
\end{figure}

Finally, client site updates are appropriately averaged to revise the global ML model parameters, $\omega$. Although a wide range of options are possible here, without loss of generality, the {\tt FedAvg} \cite{mcmahan2017},\cite{qli2021} approach is used in Figs. \ref{FL_alg} and \ref{psuedocode_server}. Namely, model averaging is done in a weighted manner based upon the size of each client site's dataset, i.e., proportional to $\frac{n_i}{N}$, where $N=\sum_i n_i$ is the aggregate amount of training data across all $K$ client sites chosen. Overall, this method is well-suited for \textit{independent identically distributed} (IID) datasets. However, researchers have also proposed modified FL server averaging schemes for heterogeneous (non-IID) local data \cite{qli2021}, and these techniques can be further integrated into the central server algorithm (left for future study).

\section{Ransomware Dataset Repository}
\label{dataset}
Realistic datasets are critical for effective ML solutions. However, as noted in Section \ref{survey}, most ransomware detection studies have at least in part, used older datasets with Windows 7 malwares. Indeed, many old ransomware control servers are no longer active as malactors have shifted to other families. In light of this, the new ransomware dataset repository curated in \cite{vehabovic2023} is used. This dataset contains binary executables of some of the most prevalent ransomware threats today, as per the IBM X-Force Threat Intelligence Index, i.e., Babuk/Babyk, BlackCat, Chaos, DJVu/STOP, Hive, LockBit, Netwalker, Sodinokibi/REvil, and WannaCry (Table \ref{table_dataset}). Namely, 9 families are chosen, and a number of Windows application executables are also collected to build a benign class and improve classifier performance. As per Fig. \ref{DRA_repository}, repository design involves two key steps, empirical data collection and feature extraction/selection, detailed next.

\subsection{Empirical Data Collection}
\label{empirical}
The work in \cite{vehabovic2023} collects a realistic set of raw binary files, with the goal of mirroring data collection at client sites (Section \ref{architecture_client}). Now the size and diversity of input data will impact ML classifier performance. For example, most algorithms yield better generalization (class separation) with larger datasets and equal representation across classes. However, given the rapidly-changing nature of ransomware, it can be difficult to obtain a sufficient number of samples for each family under consideration. Hence effective solutions must achieve good detection and attribution performance with more constrained ``minimalist'' datasets, i.e., only hundreds of samples per ransomware family. Note that other recent studies have also used smaller datasets with under 2,000 samples \cite{wang2021},\cite{zhu2022}. This requirement is well-aligned with broader trends in \textit{artificial intelligence} (AI) to develop more ``data-centric'' solutions for specialized problems \cite{ng2022}.  Consider some further details.

Malware samples are extracted from various online sites to capture some of the latest ransomware threats. Now many portals allow users to upload/download malware executables, e.g., {\tt MalwareBazar}, {\tt Triage}, {\tt VirusShare}, and {\tt VirusTotal}, etc. However, these repositories provide varying access and usability. For example, {\tt VirusTotal} and {\tt VirusShare} require user registration to access private repositories. Detailed cross-checking and analysis of samples also reveals notable duplication across portals. For example, many Sodinokibi samples on {\tt MalwareBazar} match those on {\tt Triage}.  There is also notable discrepancy between the number of samples for each family. For example DJVu is relatively abundant whereas Babuk/Babyk is more scarce and harder to find. Finally, repositories (such as {\tt VirusShare}) do not label their raw data, further complicating collection. Hence samples from these large unlabeled data dumps have to be analyzed individually using hashing programs and then cross-checked with other labelled samples (a tedious and time-consuming process). While other repositories like {\tt VirusTotal} may have labelled samples, they do not allow researchers to download them freely. Given these realities, there is potential for a lack of diversity (even scarcity) of unique samples for new ransomware families.
\begin{figure*}[h]
    \centering
    \includegraphics[width=5.0in, height=2.25in]{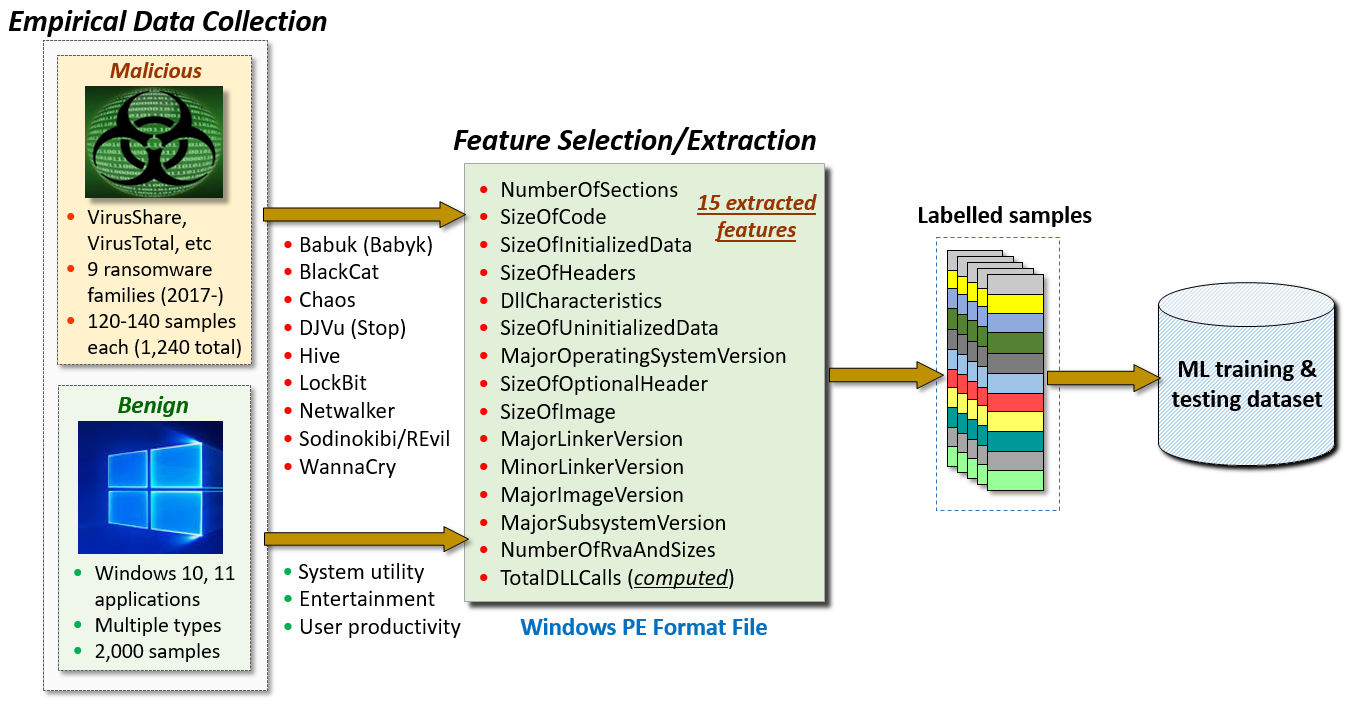}
    \caption{Ransomware repository design and static PE file features (as per \cite{vehabovic2023})}
    \label{DRA_repository}
\end{figure*}
\begin{table}[h]
	\centering
\caption{Empirical ransomware dataset (as per \cite{vehabovic2023})}
\label{table_dataset}
\begin{tabular}{|l|c|c|c|} 
\hline
\textbf{ Family } & \textbf{Samples} & \textbf{Avg. Size}  & \textbf{Avg. PE File} \\ \hline \hline
Babuk (Babyk) & 140 & 0.19 MB & 32.68 KB \\ \hline
BlackCat & 140 & 3.91 MB & 1,147 KB \\ \hline
Chaos & 140 & 0.49 MB & 35.2 KB \\ \hline
DJVu (STOP) & 140 & 0.71 MB & 66.2 KB \\ \hline
Hive & 140 & 3.51 MB & 403.9 KB \\ \hline
LockBit & 140 & 1.30 MB & 171.5 KB \\ \hline
Netwalker & 140 & 0.26 MB & 35.72 KB \\ \hline
Sodinokibi & 140 & 0.30 MB & 50.89 KB \\ \hline
WannaCry & 140 & 7.62 MB & 21.83 KB \\ \hline
\hline
Benign & 2,000 & 26.86 MB & 155.88 KB \\ \hline
\end{tabular}  
\end{table}

In light of the above, a ``minimalist'' raw dataset is collected for the 9 ransomware families \cite{vehabovic2023}. Specifically, only 140 executable samples are gathered for each family, yielding a total of 1,260 malicious samples (under 1,500 samples). However, unlike some recent studies in ransomware classification \cite{wang2021},\cite{zhu2022}, a large number of Windows 10/11 applications are also downloaded to build a benign class (2,000 in total). These programs are collected from a range of websites and include system utility, entertainment, and productivity tools (Fig. \ref{DRA_repository}). Overall, the addition of a benign class is crucial for ML purposes as it can help establish a more clear delineation between malicious programs and reduce classification errors. Further details on the collected samples are also presented in Table \ref{table_dataset}. 
\subsection{Feature Selection/Extraction}
\label{feature_extraction}
ML performance is also very dependent upon the type of input training data. It is here that feature extraction (engineering) plays a vital role in transforming raw sample files (executables) and generating meaningful information for classifiers \cite{geron2019}. Now as noted in Section \ref{survey_ransomware}, static analysis is more effective for targeting the early stages of the kill-chain. Hence this strategy is chosen here. In particular a small set of static parameters are extracted from Windows PE format files to generate a very lightweight set of features (i.e., hundreds of bytes per sample, Fig. \ref{DRA_repository}). This selection contrasts with other ransomware schemes which use much more compute-intensive methods to extract larger/heavier feature sets, e.g., using code extraction \cite{zhang2019}, image processing \cite{wang2021}, and entropy \cite{zhu2022} based methods.

Overall, PE format files are used to support executables running in 32-bit and 64-bit Windows OS settings \cite{win_PE_text} (and similar constructs also exist for other OS types). This file uses the \textit{common object file format} (COFF) and contains key information for the OS loader to run wrapped executable code. Namely, a PE file specifies memory mapping and permissions and is comprised of initial lead-in headers and multiple sections. Each section contains actual file contents (such as code or data) and has its own header \cite{win_PE_text}. Overall, PE files contain a wealth of static information, and different executables can have unique non-overlapping parameters as per functionality. As a result, it is important to extract a proper subset of parameters here, i.e., feature vectors. Foremost, a chosen feature should exist in the PE format files for all executables in the repository. Additionally, selected parameter ranges should exhibit sufficient variability across classes.

In light of the above, PE files are generated for all raw executable files. Careful experimentation is then done to select a total of 15 PE file parameters and construct labelled feature vectors to build the aggregate ML training/testing dataset, Fig. \ref{DRA_repository} (i.e., 1,260 ransomware vectors and 2,000 benign vectors). Only entries from the \textit{Image\_File\_Header}, \textit{Image\_Optional\_Header}, and \textit{Image\_Section\_Header} sections are chosen here, along the lines of earlier studies in \cite{kim2016}-\cite{rezaei2020}. Some selected PE parameters include \textit{NumberOfSections}, \textit{SizeOfCode}, \textit{SizeOfHeaders}, etc. Note that PE format files also contain information on \textit{dynamic-link library} (DLL) calls, and this can shed light on program functionality. For example, most ransomware programs use encryption, socket-based communication, and registry-modification functions. Hence the total number of DLL calls is also added to the feature vector, i.e., \textit{TotalDLLCalls} (Fig. \ref{DRA_repository}). Note that this is a \textit{computed} feature and not a parameter read in from the PE file. 

\section{Performance Evaluation}
\label{performance}
As detailed in Section \ref{architecture}, the DRA architecture represents a generic approach which can implement a wide range of (NN-based) ransomware analysis schemes. Hence this framework is evaluated for the case of static analysis using the ML training/testing dataset curated in Section \ref{dataset}. All model development and testing is done using the {\tt Keras} and {\tt TensorFlow} libraries, along with other Python-based toolkits such as {\tt Pandas}, {\tt Numpy}, and {\tt Sklearn}. Complete details on the testing setup and results from the performance evaluation study are now presented.

\begin{table}[]
\centering
\caption{Aggregate dataset partitioning (1,260 ransomware and 2,000 benign samples)}
\begin{tabular}{|c|l|clccc|l|cc|l|cl|}
\hline
\multicolumn{1}{|l|}{\multirow{3}{*}{\textbf{ Clients }}} & \multirow{3}{*}{} & \multicolumn{5}{c|}{\textbf{Ransomware}}   & \multirow{3}{*}{} & \multicolumn{2}{c|}{\textbf{Benign}}                       & \multirow{3}{*}{} & \multicolumn{2}{c|}{\multirow{3}{*}{\textbf{Malware \%}}} \\ \cline{3-7} \cline{9-10}
\multicolumn{1}{|l|}{}   &   & \multicolumn{3}{c|}{\textbf{Training}}  & \multicolumn{2}{c|}{\textbf{Testing}}   &  & \multicolumn{1}{c|}{\multirow{2}{*}{\textbf{Training}}} & \multirow{2}{*}{\textbf{Testing}} &  & \multicolumn{2}{c|}{} \\ \cline{3-7}
\multicolumn{1}{|l|}{}  &  & \multicolumn{2}{c|}{ Per Family } & \multicolumn{1}{c|}{ Total }   & \multicolumn{1}{c|}{ Per Family } & { Total } &  & \multicolumn{1}{c|}{}                                   &  &  & \multicolumn{2}{c|}{}  \\ \hline \hline
$K$=1 &   & \multicolumn{2}{c|}{120}  & \multicolumn{1}{c|}{1,080}     & \multicolumn{1}{c|}{20}  & 180   &  & \multicolumn{1}{c|}{1,700}   & 300   &   & \multicolumn{2}{c|}{38\%}                                 \\ \hline
$K$=2  &  & \multicolumn{2}{c|}{60}  & \multicolumn{1}{c|}{540}     & \multicolumn{1}{c|}{20}  & 180   &   & \multicolumn{1}{c|}{1,700}  & 300  &   & \multicolumn{2}{c|}{24\%}                                 \\ \hline
$K$=3   &    & \multicolumn{2}{c|}{40}  & \multicolumn{1}{c|}{360}     & \multicolumn{1}{c|}{20} & 180   &  & \multicolumn{1}{c|}{1,700}    & 300   &   & \multicolumn{2}{c|}{17\%} \\ \hline
$K$=4  &   & \multicolumn{2}{c|}{30}  & \multicolumn{1}{c|}{270}  & \multicolumn{1}{c|}{20} & 180   &     & \multicolumn{1}{c|}{1,700} & 300   &   & \multicolumn{2}{c|}{14\%}                                 \\ \hline
\end{tabular}
\label{table_dataset_split}
\end{table}

FL performance is tested for a varying number of client sites ($K$=2, 3, 4) by partitioning the ML training/testing dataset, see Table \ref{table_dataset_split}. Recall that this aggregate dataset contains feature vectors (samples) for each raw binary file (where each feature vector contains 15 extracted PE  file parameters, Fig. \ref{DRA_repository}). Now there are 140 samples per family and 2,000 benign applications, i.e., total of 10 classes. Hence 120 samples from each ransomware family are randomly selected for training (1,080 total) and the remaining 20 are used for testing (180 total). This partitioning represents a 85/15\% split between training and testing data. Next, the 120 ransomware samples (per class) are further partitioned between the client sites, yielding several local ML training datasets, i.e., full 120 samples per class for $K$=1 (centralized ML), 60 samples per class for $K$=2 client sites (540 total), 40 samples per class for $K$=3 client sites (360 total), and 30 samples per class for $K$=4 client sites (270 total), see Table \ref{table_dataset_split}. Meanwhile the benign dataset is not partitioned between the client sites, thereby yielding a higher percentage of non-malicious training data. This choice mirrors realistic settings where regular applications downloads will exceed ransomware downloads. Moreover, there may be high commonality between application downloads across organizations, and hence it is plausible to use the same set of benign samples across client sites. Accordingly, the benign samples are split in a 85/15\% manner, with 1,700 samples randomly selected for training and the remaining 300 for testing (per client site). Overall, the proportion of ransomware training data declines with the number of client sites, i.e., malware percentage, Table \ref{table_dataset_split}. For example, with 4 client sites there is more than 6 times less training data (i.e., 270 vs 1,700 samples).  
Carefully note that in practice, different regions (client sites) will experience varying types of ransomware attacks.  These differences will yield unbalanced datasets with possibly different underlying distributions, and are left for future study.


Ransomware analysis is conducted using a range of supervised ML classifiers. These algorithms are trained and tested for the classification/attribution problem with 10 classes, i.e., 9 ransomware, 1 benign (Section \ref{dataset}). Foremost, the FL approach uses the supervised FNN algorithm (with gradient descent) to train local ML models at client sites. This network has 2 hidden layers and 32 nodes per layer, yielding a relatively small set of parameter weights (i.e., $\omega$, Section \ref{architecture}). As noted earlier, FL can be used with many NN variants, but the FNN algorithm is chosen here given the smaller dataset sizes and feature vectors involved. Client sites train their local models over $E$ epochs and send model parameter updates to the central server. Meanwhile, global training is done over $T$ rounds. Several centralized ML algorithms are also used for comparison purposes, including centralized FNN ($K$=1), \textit{support vector machines} (SVM), and \textit{random forest} (RF) \cite{geron2019}. These classifiers are evaluated using the complete dataset with a 85/15\% partitioning between training/testing samples. All results (FL, centralized) are averaged over 50 independent trails, with each using different randomized selections of the requisite dataset partitions. The control parameters for different ML algorithms are also fine-tuned to achieve the best classification rates.

Several metrics are used to gauge the ML classifiers, including accuracy, precision, recall, and F1 score. For example, precision is a measure of correctness whereas recall is a measure of relevance \cite{geron2019}. Two additional metrics are also defined to capture the \textit{binary} detection capabilities of multi-class classifiers, i.e., selection between ransomware and benign. Consider mis-classification in more detail. Here, incorrectly classifying ransomware as benign is much more problematic than mis-classifying it as another family (as it may bypass network or host defenses). To quantify this, a \textit{ransomware detection rate} (RDR) is defined:
\begin{equation}
    RDR = \frac{T_{rs}}{T_{rs}+F_{rs}}
    \label{eq_RDR}
\end{equation}
where $T_{rs}$ is the total number of ransomware testing samples across all families that are classified as (any family of) ransomware, and $F_{rs}$ is the total number of ransomware testing samples across all families that are mis-classified as benign, i.e., total number of ransomware testing samples is $(T_{rs}+F_{rs})$.  This metric is similar to recall and treats all ransomware families as a single malicious class, i.e., tracks false negatives. Similarly, a \textit{benign detection rate} (BDR) is also defined:
\begin{equation}
     BDR = \frac{T_{bn}}{T_{bn}+F_{bn}}
    \label{eq_BDR}
\end{equation}
where $T_{bn}$ is the total number of benign testing samples classified as benign, and $F_{bn}$ is the total number of benign testing samples mis-classified as ransomware, i.e., total number of benign testing samples is $(T_{bn}+F_{bn})$. Note that the false negative classification of benign applications is generally less of a security threat than false negative classification of ransomware programs.  

\begin{table}[h]
	\centering
\caption{Results for FL with varying client sites, $K$ ({\tt FedAvg})}
\label{table_results1}
\begin{tabular}{|c||c|c|c|c|} 
\hline
\textbf{2 Client Sites } & \textbf{ Accuracy } & \textbf{ Precision }  & \textbf{ Recall } & \textbf{ F1 Score } \\ \hline \hline
{ Global Avg. Model } & \textbf{91.87}\% & \textbf{86.01}\% & \textbf{86.67}\% & \textbf{91.82}\% \\ \hline  
\hspace{0.25in} Client site 1 & 86.52\% & 78.57\% & 79.37\% & 86.30\% \\ \hline
\hspace{0.25in} Client site 2 & 88.34\% & 78.59\% & 79.22\% & 87.34\% \\ \hline
\textbf{ 3 Client Sites } & \textbf{ Accuracy } & \textbf{ Precision }  & \textbf{ Recall } & \textbf{ F1 Score } \\ \hline \hline
{ Global Avg. Model } & \textbf{92.11\%} & \textbf{87.40\%} & \textbf{86.57\%} & \textbf{91.90\%} \\ \hline  
\hspace{0.25in} Client site 1 & 86.02\% & 77.86\% & 75.64\% & 85.33\% \\ \hline
\hspace{0.25in} Client site 2 & 82.90\% & 73.89\% & 74.56\% & 82.13\% \\ \hline
\hspace{0.25in} Client site 3 & 81.47\% & 73.55\% & 75.95\% & 81.49\% \\ \hline
\textbf{ 4 Client Sites } & \textbf{ Accuracy } & \textbf{ Precision }  & \textbf{ Recall } & \textbf{ F1 Score } \\ \hline \hline
{ Global Avg. Model }  & \textbf{92.46\%} & \textbf{87.69\%} & \textbf{86.40\%} & \textbf{92.43\%} \\ \hline   
\hspace{0.25in} Client site 1 & 79.93\% & 72.22\% & 72.61\% & 79.53\% \\ \hline
\hspace{0.25in} Client site 2 & 83.70\% & 73.29\% & 74.41\% & 83.48\% \\ \hline
\hspace{0.25in} Client site 3 & 85.48\% & 71.91\% & 71.73\% & 83.76\% \\ \hline
\hspace{0.25in} Client site 4 & 85.49\% & 72.23\% & 71.41\% & 80.20\% \\ \hline
\end{tabular}  
\end{table}

Detailed results for the FL scheme are first presented in Table \ref{table_results1} for varying numbers of client sites ($K$ values). In each case, the table lists the accuracy, precision, recall, and F1 scores for the global averaged FNN model (in bold) followed by the individual local client site models. These findings show vastly improved model generalization with FL, with the global models exceeding the individual client models by sizeable margins for all metrics. For example, average accuracy is 3-12\% higher, precision is 7-15\% higher, recall is 7-15\% higher, and F1 scores are 4-13\% higher. These are very impressive results and indicate that organizations can greatly improve their ransomware defenses by participating in FL-based schemes, i.e., while retaining data privacy and maintaining smaller datasets. Furthermore, FL accuracy is also very high. For example, the accuracy (and F1 score) with 4 client sites is close to 92.5\%, i.e., correct attribution of over 18 out of 20 samples. These results also closely match some centralized ransomware classification schemes (Section \ref{survey_ransomware}) many of which implement heavier feature extraction and ML computation algorithms, e.g., image and entropy-based features, deep NN designs, etc \cite{wang2021},\cite{zhu2022}.  By contrast, the FL setup herein uses very small feature vectors (15 parameters) and basic FNN models.

\begin{table}[]
\centering
\caption{Results for ML models (averaged over 50 independent runs)}
\label{table_results2}
\begin{tabular}{|c||cl|cl|cl|c||cc|}
\hline
\multicolumn{1}{|c|}{\multirow{2}{*}{\textbf{ Distd. FL}}} & \multicolumn{2}{c|}{\multirow{2}{*}{\textbf{ Accuracy }}} & \multicolumn{2}{c|}{\multirow{2}{*}{\textbf{ Precision }}} & \multicolumn{2}{c|}{\multirow{2}{*}{\textbf{ Recall }}} & \multirow{2}{*}{\textbf{ F1 Score }} & \multicolumn{2}{c|}{\textbf{ Binary }}   \\ \cline{9-10} 
\multicolumn{1}{|c|}{}                                         & \multicolumn{2}{c|}{}                                   & \multicolumn{2}{c|}{}                                    & \multicolumn{2}{c|}{}                                 &                                    & \multicolumn{1}{c|}{\textbf{ RDR }} & \textbf{ BDR } \\ \hline
2 clients (FNN)                                               & \multicolumn{2}{c|}{ 91.87\%}                              & \multicolumn{2}{c|}{ 86.01\%}                               & \multicolumn{2}{c|}{ 86.67\%}                            &  { 91.82\%}                              & \multicolumn{1}{c|}{ 94.92\% }        & { 94.99\%  }       \\ \hline
 3 clients (FNN)                                               & \multicolumn{2}{c|}{ 92.11\%}                              & \multicolumn{2}{c|}{ 87.40\%}                               & \multicolumn{2}{c|}{ 86.57\%}                            &  { 91.90\%}                              & \multicolumn{1}{c|}{ 93.04\% }        & { 95.86\% }        \\ \hline
4 clients (FNN)                                               & \multicolumn{2}{c|}{ 92.46\%}                              & \multicolumn{2}{c|}{ 87.69\%}                               & \multicolumn{2}{c|}{ 86.40\%}                            & { 92.43\%}                              & \multicolumn{1}{c|}{ 93.27\% }        & { 96.43\% }        \\ \hline
\multicolumn{1}{|c|}{\multirow{2}{*}{\textbf{ Centralized }}}    & \multicolumn{2}{c|}{\multirow{2}{*}{\textbf{ Accuracy }}} & \multicolumn{2}{c|}{\multirow{2}{*}{\textbf{ Precision }}} & \multicolumn{2}{c|}{\multirow{2}{*}{\textbf{ Recall }}} & \multirow{2}{*}{\textbf{ F1 Score }} & \multicolumn{2}{c|}{\textbf{ Binary }}   \\ \cline{9-10} 
\multicolumn{1}{|c|}{}                                         & \multicolumn{2}{c|}{}                                   & \multicolumn{2}{c|}{}                                    & \multicolumn{2}{c|}{}                                 &                                    & \multicolumn{1}{c|}{\textbf{ RDR }} & \textbf{ BDR } \\ \hline \hline
FNN                                                            & \multicolumn{2}{c|}{91.48\%}                              & \multicolumn{2}{c|}{86.84\%}                               & \multicolumn{2}{c|}{84.68\%}                            & 91.27\%                              & \multicolumn{1}{c|}{ 92.06\% }        & 96.69\%        \\ \hline \hline
SVM                                                            & \multicolumn{2}{c|}{90.44\%}                              & \multicolumn{2}{c|}{91.16\%}                               & \multicolumn{2}{c|}{75.60\%}                            & 89.86\%                              & \multicolumn{1}{c|}{ 80.39\% }        & { 97.85\%  }       \\ \hline
RF                                                             & \multicolumn{2}{c|}{96.02\%}                              & \multicolumn{2}{c|}{94.41\%}                               & \multicolumn{2}{c|}{92.07\%}                            & 95.98\%                              & \multicolumn{1}{c|}{ 95.72\% }        & 99.05\%        \\ \hline
\end{tabular}
\end{table}

Next, the performance of all ML classifiers is presented in Table \ref{table_results2}. 
Here the respective FL percentages are the same as the global model averages from Table \ref{table_results1}. First consider multi-class attribution, as measured by the accuracy, precision, recall, and F1 scores. Foremost, the FL approach outperforms its centralized FNN counterpart for varying numbers of client sites. In particular with $K$=4 client sites the accuracy and F1 scores are 1\% higher. These gains come despite using much smaller training datasets at the client sites, e.g., only 270 ransomware samples for $K$=4, Table \ref{table_dataset_split}. Again, this is another key result as it demonstrates the ability of decentralized FL setups (with smaller client sites) to achieve similar or better ransomware attribution compared to less practical centralized setups (requiring much more training data). Furthermore, the FL approach also outperforms the SVM algorithm by over 2\% in terms of accuracy. However, the centralized RF scheme (trained with global data) gives the best results, with accuracy and F1 scores averaging about 3.5\% higher than FL.
\begin{figure*}[h]
    \centering
    \includegraphics[width=4.5in, height=3.25in]{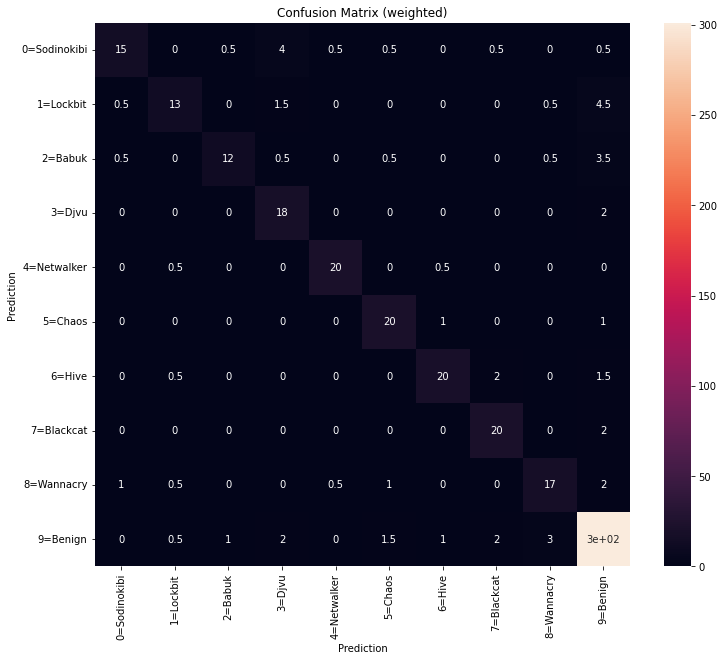}
    \caption{Sample multi-class confusion matrix for FL ($K$=4 client sites, $E$=1 epoch)}
    \label{fig_confusion_matrix}
\end{figure*}

Also, Table \ref{table_results2} presents \textit{binary} detection results, as measured by the RDR and BDR metrics (Eqs. \ref{eq_RDR}, \ref{eq_BDR}). The former is deemed more important as it captures the mis-classification rate of ransomware. Overall, FL gives very good RDR values, up to 94.92\%, and within 1\% of the RF scheme which has the best binary results. Namely, this scheme yields a false negative rate of about 1 in 20 malicious samples. These results are impressive and match those from earlier studies on binary detection of older ransomware threats, Section \ref{survey}. Also, the BDR values are higher than the RDR values since training datasets have a larger proportion of benign data as per real-world settings (Table \ref{table_dataset_split}). Note that the FNN-based schemes (including distributed FL) give slightly lower BDR values than the other ML algorithms, i.e., by about 1-2.5\%. Nevertheless, the related BDR percentages are still close to 97\%, i.e., 1 error in about 33 benign samples.

\begin{figure*}[h]
    \centering
    \includegraphics[width=3.65in, height=1.90in]{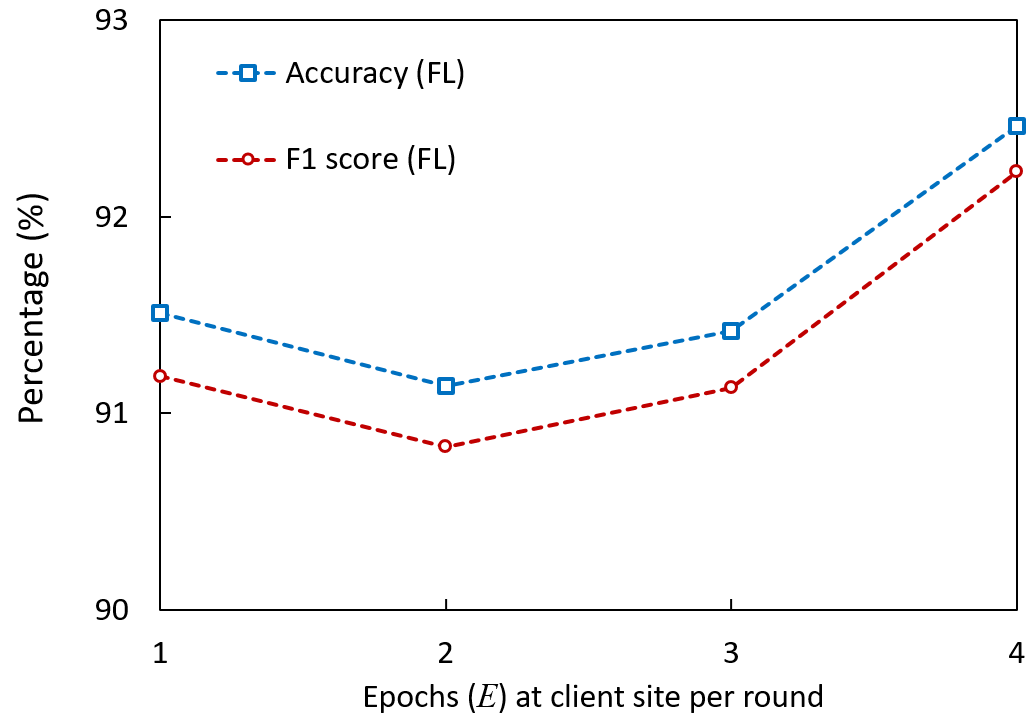}
    \caption{Accuracy and F1 score results for $K$=4 client sites (averaged over 50 trials)}
    \label{fig_accuracy_F1}
\end{figure*}

To investigate detailed per-class behaviors, Fig. \ref{fig_confusion_matrix} shows a sample averaged confusion matrix for FL with 4 client sites (rows 0-9 represent ransomware families and row 10 represents benign applications). Note that the numbers in row 10 are larger as there are more benign testing samples, Table \ref{table_dataset_split}. Here, the majority of ransomware and benign application samples are correctly classified, i.e., diagonal entries dominate. Furthermore, most mis-attributed ransomware samples are still classified as some form of ransomware, although LockBit and Babuk show higher averages, i.e., 4.5 (22.5\%) and 3.5 (17.5\%) out of 20, respectively. Hence the potential damage from false negatives is relatively small for most ransomware cases. Meanwhile, Fig. \ref{fig_accuracy_F1} plots the average accuracy and F1 scores for the global FL model with $K$=4 client sites and varying epoch counts ($E$). These results shed light on the local FL training process and indicate better performance with $E$=4 epochs (also observed for several other $K$ values).
\begin{figure*}[h]
    \centering
    \includegraphics[width=4.55in, height=1.90in]{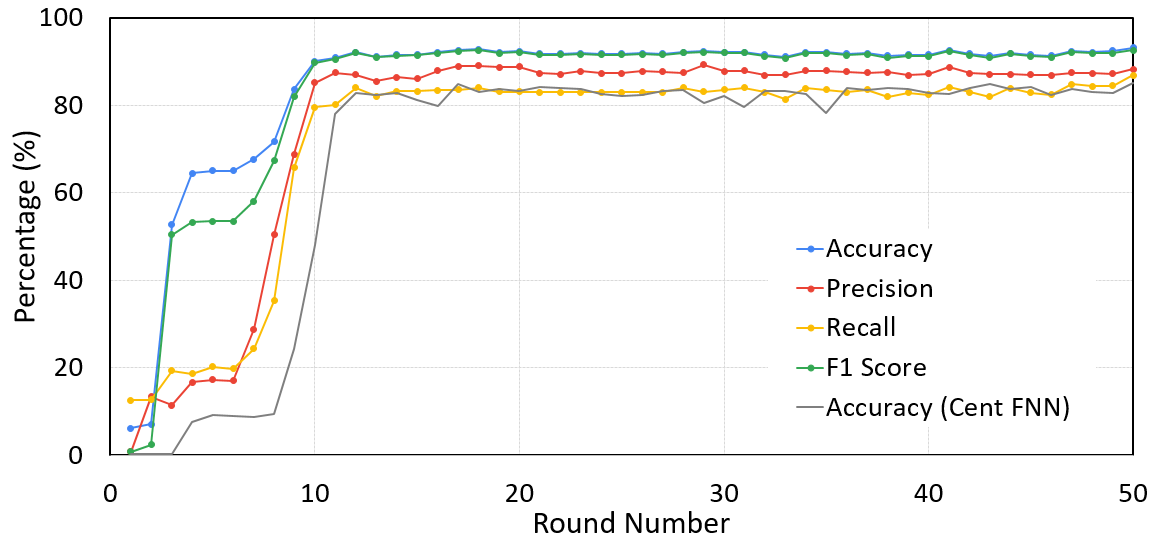}
    \caption{ Sample progression of FL training ({\tt FedAvg}, $K$=4 client sites, $E$=4 epochs)}
    \label{fig_rounds}
\end{figure*}

Finally, Fig. \ref{fig_rounds} plots the accuracy, precision, recall, and F1 scores at the end of each communication round for the global FL model with $K$=4 client sites (and $E$=4 epochs). For reference sake, corresponding accuracy values for the centralized FNN classifier are also plotted. The objective here is to observe the FL training process over multiple rounds. These plots show very rapid progression, with performance improving rapidly within the first 10 rounds and stabilizing by about 15 rounds. By contrast, the centralized FNN tales more rounds to improve, with accuracy picking up after 9 rounds. Overall, these results demonstrate improved responsiveness and learning with the distributed FL approach, further complementing its inherent privacy and scalability benefits.

\section{Conclusions \& Future Work}
\label{conclusions}
Ransomware presents a persistent threat to users and organizations. 
Researchers have developed various solutions to detect and classify this malware using \textit{machine learning} (ML) schemes to analyze binary files and system/network activities. However, growing scalability and privacy concerns make it difficult to collect and analyze massive amounts of data at a centralized site. Hence this study presents a novel \textit{distributed ransomware analysis} (DRA) framework for detection and attribution using \textit{federated learning} (FL). This architecture embodies a generic approach which can implement both static and dynamic analysis schemes. A realistic dataset repository comprising of some of the latest ransomware threats is used to conduct a detailed performance evaluation study for the case of static analysis with rapid/lightweight feature extraction from Windows \textit{portable executable} (PE) format files. Overall findings confirm superior performance for the FL-based approach, with global ML model performance notably exceeding locally trained models. The FL approach also closely matches or outperforms some centralized ML algorithms in terms of attribution accuracy and binary detection. 


Overall, this effort presents one of the first studies on ransomware detection and attribution using the decentralized FL framework and provides a strong basis for further study. Foremost, a broader set of static and dynamic features can be used to improve local model training (leveraging existing work on ransomware detection). Further efforts can also address FL bias and variability concerns.

\section{Acknowledgements}
This work has been supported in part by Cyber Florida.  The authors are very grateful for this support.

%
%
%
\bibliographystyle{IEEEtran}
\bibliography{references.bib}

\end{document}